%% file: Paparella.Labee.ea.ITSC24.tex
\documentclass[letterpaper, 10 pt, conference]{ieeeconf} 
\pdfoutput=1 
\IEEEoverridecommandlockouts
\usepackage[english]{babel}
\usepackage{cite}
\usepackage{amsmath,amssymb,amsfonts}
\usepackage{algorithm}
\usepackage{algorithmic}
\usepackage{graphicx}
\usepackage{textcomp}
\usepackage{amsmath}
\usepackage{dsfont}
\usepackage{xcolor}
\usepackage{units}
\usepackage{booktabs}
\usepackage{comment}
\usepackage{url}
\usepackage[normalem]{ulem}
\usepackage[hidelinks]{hyperref}
\usepackage{caption}
\usepackage{subcaption}
\usepackage{flushend}
\usepackage{lettrine}
\usepackage{eurosym}

\usepackage{amsthm}
\newcounter{thm}
\newtheorem{prob}[thm]{Problem}

\newtheorem{theorem}{Theorem}[section]

\def\BibTeX{{\rm B\kern-.05em{\sc i\kern-.025em b}\kern-.08em
		T\kern-.1667em\lower.7ex\hbox{E}\kern-.125emX}}

\newif\ifmargincomments
\margincommentstrue 

\newif\ifrev
\revtrue 

\input{def}

\begin{document}

\title{\LARGE \bf Multi-objective Optimal Trade-off Between V2G Activities and Battery Degradation in Electric Mobility-as-a-Service Systems}

\author{Fabio Paparella, Pim Labee, Steven Wilkins, Theo Hofman, Soora Rasouli, Mauro Salazar
\thanks{This publication is part of the project NEON with number 17628 of the research program Crossover, partly financed by the Dutch Research Council. The authors are with the Control Systems Techonology section, Eindhoven University of Technology, Eindhoven, The Netherlands
        {\tt\small \{f.paparella, p.labee, swilkins, t.hofman, s.rasouli,  m.r.u.salazar\}@tue.nl}.      
    }%
}

\maketitle
\thispagestyle{empty}
\pagestyle{empty}

\begin{abstract}
This paper presents optimization models for electric Mobility-as-a-Service systems, whereby electric vehicles not only provide on-demand mobility, but also perform charging and Vehicle-to-Grid (V2G) operations to enhance the fleet operator profitability.
Specifically, we formulate the optimal fleet operation problem as a mixed-integer linear program, with the objective combining of operational costs and revenues generated from servicing requests and grid electricity sales.
Our cost function explicitly captures battery price and degradation, reflecting their impact on the fleet total cost of ownership due to additional charging and discharging activities.
Simulation results for Eindhoven, The Netherlands, show that integrating V2G activities does not compromise the number of travel requests being served.
Moreover, we emphasize the significance of accounting for battery degradation, as the costs associated with it can potentially outweigh the revenues stemming from V2G operations.
\end{abstract}
\input{Sections/Introduction.tex}
\input{Sections/Problem.tex}
\input{Sections/CaseStudy.tex}
\input{Sections/Conclusions.tex}

\bibliographystyle{IEEEtran}
\bibliography{main.bib,SML_papers.bib,mobility.bib}

\end{document}

%% file: def.tex
\ifmargincomments

\else

\fi

\ifrev

\else

\fi

\newcommand{\flexbrac}[1]{\if\relax\detokenize{#1}\relax \else (#1) \fi}
\newcommand{\flexcomma}[1]{\if\relax\detokenize{#1}\relax \else ,#1 \fi}

\newcommand{\abs}[1]{\lvert #1 \rvert}

\newcommand{\cC}{\mathcal{C}}

\newcommand{\cG}{\mathcal{G}}

\newcommand{\cI}{\mathcal{I}}

\newcommand{\cK}{\mathcal{K}}

\newcommand{\cR}{\mathcal{R}}



%% file: Sections/Introduction.tex
\section{Introduction}
\begin{figure}[t]
	\centering
	\includegraphics[trim={0 0 180 0},clip,width=0.7\columnwidth]{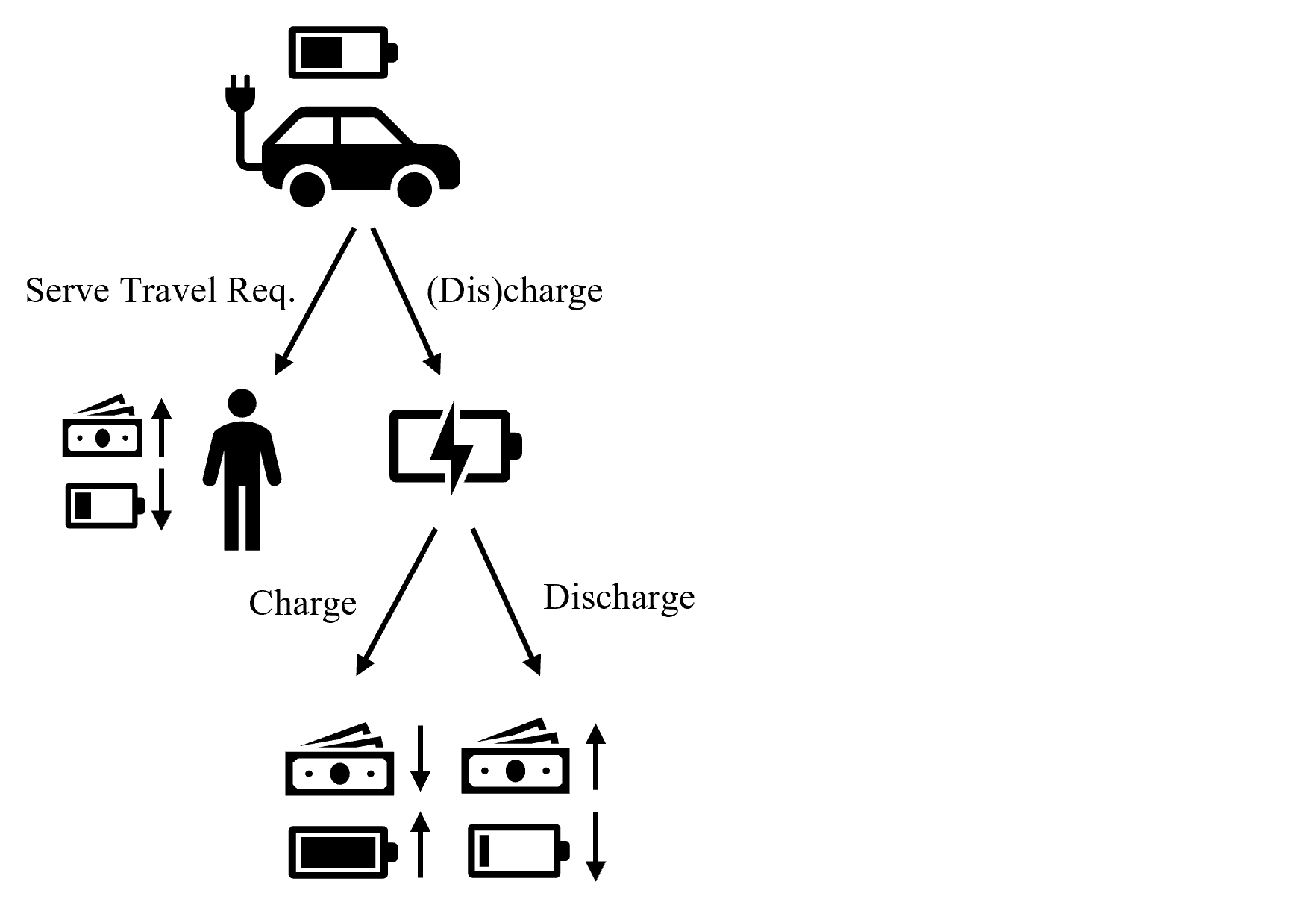}
	\caption{Schematic representation of the decision tree of one vehicle. At every step, each electric vehicle can either serve a travel requests, charge or discharge the battery. } 
	\label{fig:1}
\end{figure}

\lettrine{R}{ising} Mobility-as-a-Service (MaaS) companies are deeply changing transportation systems by providing on-demand mobility to users through fleets of vehicles. However, the whole transportation sector is still facing an increase in emissions and traffic congestion. In addition, when dealing with electric vehicles, a new problem related to congestion of electric grid arises. In particular, electric grid instability and congestion is caused by a spatial-temporal mismatch between supply and demand which, in turn, can cause strong volatility in electricity prices. Electric vehicles can be seen as decentralized energy storage and have the potential to leverage such volatility of electricity prices to improve the profitability of the fleet operator by buying and selling while they are not providing on-demand mobility, see Fig.~\ref{fig:1}. At the same time, however, battery degradation shall be taken into account for when studying the potential of this solution.
This paper presents an optimization framework for a fleet of autonomous electric vehicles that provide on-demand mobility and can perform charging and vehicle-to-grid (V2G) activities, whilst accounting for battery degradation. 

\textit{Related Literature:}
Mobility-on-demand systems have been extensively studied leveraging many different methods~\cite{PavoneFrazzoli2010,TsaoIglesiasEtAl2018,PaparellaPedrosoEtAl2024b,PaparellaPedrosoEtAl2024,Wollenstein-BetechHoushmandEtAl2020,RossiZhangEtAl2017,VehlhaberSalazar2023b,WallarAlonso-MoraEtAl2019,VakayilGruelEtAl2017,HoerlRuchEtAl2018b,FiedlerCertickyEtAl2018,ChenLowEtAl2015,PedrosoHeemelsEtAl2023}.
In particular, the ones dealing with electric fleets are numerous, mainly focusing on the charging placement and sizing, and on the interaction with power networks ~\cite{LukeSalazarEtAl2021,RossiIglesiasEtAl2018b,AlizadehWaiEtAl2014,AlizadehWaiEtAl2016}. The authors of~\cite{LeFlochMeglioEtAl2015} study the optimal aggregate charging and V2G operation of a fleet of vehicles with fixed time slots.
However, none of these studies include battery degradation. In~\cite{WangCoignardEtAl2016,UddinDubarryEtAl2018} the authors study the trade-off between battery degradation and V2G for a private vehicle, while in~\cite{RathMedinaEtAl2023} the authors study the same trade-off for a fleet of electric vehicles for mobility services, but they assume to know the fixed time-slots when it is possible to charge, namely the mobility-on-demand problem is decoupled from the (dis)charging activities.
To conclude, to the authors' knowledge, no model that investigates electric mobility-on-demand systems whilst accounting for battery degradation has been proposed yet.

\textit{Statement of Contributions:}
This work presents an optimization framework to analyze the mobility operations of a fleet of electric vehicles for MaaS, while modeling charging and V2G activities, taking into account battery degradation, whereby the objective is to maximize the profitability of the fleet operator. In particular, we devise a linear approximation of a battery degradation model that allows to write the full problem in a mixed integer linear fashion that can be handled by commercial solvers.


%% file: Sections/Problem.tex
\section{Optimization Framework}\label{sec:Form}
\begin{figure}[t]
	\centering
	\includegraphics[trim={50 0 40 10},clip,width=\columnwidth]{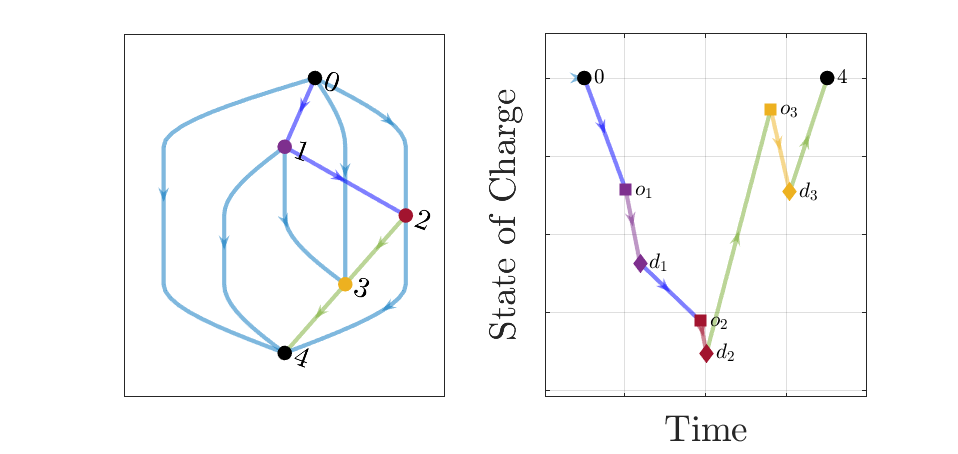}
	\caption{Directed Acyclic Graph (DAG) representation showing the operation of one vehicle. Figure taken from~\cite{PaparellaHofmanEtAl2024}.} 
	\label{fig:DAG}
\end{figure}

In this section, we show the mobility-on-demand problem coupled with the charging and V2G operations, and battery degradation, which is partially based on our previous work~\cite{PaparellaHofmanEtAl2024}, Fig.~\ref{fig:DAG}. 
First, we define a set of travel requests $i \in \mathcal{I}:=\{1,2...,I\}$ as $r_i = (o_i,d_i,t_i)$ being a user that asks for a point-to-point mobility service, namely being picked up from an origin $o_i$ to a destination $d_i$ at time $t_i$. 
In particular, we expand the set of travel requests to $\mathcal{I}^+:=\{0,1,2...,I,I+1\}$ which allows to represent the initial and final position of the fleet at a pre-defined location. We construct a Directed Acyclic Graph (DAG), $\mathcal{G}=(\cI^+,\mathcal{A})$, where arc $(i,j) \in \mathcal{A}$ is the fastest path from the destination $d_i$ to the origin $o_j$. The travel time and distance are $t^\mathrm{fp}_{ij},d^\mathrm{fp}_{ij}$, respectively. Note that for simplicity we indicate with $t^\mathrm{fp}_{ii},d^\mathrm{fp}_{ii}$ the time and distance of the fastest path to serve $r_i$.
Last, we define a set of vehicles $\cK :=\{1,2...,K\}$ and a set of charging stations $\mathcal{C} :=\{1,2...,C\}$.
First, we associate a variable to each arc through a transition tensor \mbox{$X \in \{0,1\}^{|\cI^+|\times |\cI^+| \times |\cK|}$}, where the entry $X_{ij}^k$ indicates if vehicle $k$ performs transition $ij$ or not, where we define as transition $ij$ the act of serving request $i$, travel from $d_i$ to $o_j$, and be ready to serve request $j$. In the same way we define a charging tensor $S \in \{0,1\}^{|\cI^+| \times |\cI^+| \times|\cK| \times|\cC|}$ to account if vehicle $k$ goes to charging station $c$ in between transition $ij$. In addition, to quantify the amount of energy exchanged with charging station $c$, we define the tensor $C \in \cR^{|\cI^+| \times |\cI^+| \times|\cK| \times|\cC|}$. Note that, if an entry is positive, vehicle $k$ recharges its battery at station $c$ in between transition $ij$, while if it is negative, it performs V2G.   
To obtain the acyclic feature of the network $\cG$, we impose time constraints on transitions, i.e., the tensors $X,S,C$.
First, we define maximum time window length of transition $ij$ as $t^\mathrm{ava}_{ij} := t_j-t_i + t_{ii}^\mathrm{fp}$. Following this we can pre-compute the upper bounds on each tensor:
\begin{equation}\label{eq:boundX}
	X_{ij}^k\leq 
	\begin{cases}
		1 & \text{if} \; t_{ij}^{\mathrm{fp}} \leq t^\mathrm{ava}_{ij}\\
		0 & \text{otherwise},
	\end{cases}  
\end{equation}
\begin{equation}
	{S}_{ijc}^k \leq 
	\begin{cases}
		1 & \text{if} \; t_{ij}^\mathrm{fp} +  \Delta T_{ijc}^\mathrm{go2S} \leq t^\mathrm{ava}_{ij}\\
		0 & \text{otherwise},
	\end{cases}   
\end{equation}
\begin{equation}\label{eq:Cmax}
	\abs {{C}_{ijc}^k } \leq
	\begin{cases}
		\hat{C}_{ijc}^k & \text{if} \; t_{ij}^\mathrm{fp} +  \Delta T_{ijc}^\mathrm{go2S} \leq t^\mathrm{ava}_{ij}\\
		0 & \text{otherwise},
	\end{cases}   
\end{equation}
where $ \Delta T_{ijc}^\mathrm{go2S}$ is the additional time required to go to charging station $c$ during transition $ij$. Element  \mbox{$\hat{C}_{ijc}^k=( t^\mathrm{ava}_{ij} -t_{ij}^\mathrm{fp} -  \Delta T_{ijc}^\mathrm{go2S})P_\mathrm{ch}$} is the maximum amount of energy that can be exchanged at charging power $P_\mathrm{ch}$ during $ij$.
To enforce serving of the travel requests and continuity of schedules for every vehicle, we enforce
\begin{equation}\label{eq:maxonce1}
	\sum_{i\in \cI^+,k\in \cK}  X_{ij}^k \leq 1\;\;\;\;\forall j\in \cI,
\end{equation} 
\begin{equation}\label{eq:maxonce2}
	\sum_{j\in \cI,k\in \cK}  X_{ij}^k \leq 1 \;\;\;\;\forall i\in \cI^+,
\end{equation} 
\begin{equation}\label{eq:conti}
	\sum_{i\in \cI^+} X_{ij}^k - \sum_{l\in \cI^+} X_{jl}^k = 0\;\;\;\;\forall j\in \cI,\;\forall k\in \cK.
\end{equation}
Then, we guarantee that a vehicle can charge if and only if it travels to a charging station, spending additional energy:
\begin{equation}\label{eq:SOnlyIfX}
	\sum_{c\in \cC}S_{ijc}^k \leq  X_{ij}^k \;\;\;\;\forall i,j\in \cI^+,\;\forall k\in \cK,
\end{equation}
\begin{equation}\label{COnlyIfS}
	\abs{C^k_{ijc}} \leq \hat{C}_{ijc}^k  \cdot S_{ijc}^k\;\;\;\;\forall i,j\in \cI^+,\;\forall c\in \cC,\;\forall k\in \cK.
\end{equation}
\begin{equation}\label{eq:EbalTrans}
	E^k_{ij} = 
	\begin{cases}
		d^\mathrm{fp}_{ij}\cdot \bar{E}_\mathrm{con} \quad \forall i,j \in \cI^+,\forall k \in\cK |\sum_{c\in \cC} S_{ijc}^k=0
		
		\\
		\begin{split}
			(d^\mathrm{fp}_{ij} + \Delta& d_{ijc}^\mathrm{go2S} ) \cdot \bar{E}_\mathrm{con} \\
			&\forall i,j \in \cI^+,\forall c \in\cC,\forall k \in\cK |S_{ijc}^k=1,
		\end{split}
	\end{cases}
\end{equation}
\begin{align}
	e^k_{j} =  e^k_{i} - E_{ij}^k &+ \sum_{c\in \cC} C_{ijc}^k \nonumber \\ 
	&\forall i,j \in \cI^+,\forall k \in \cK |X_{ij}^k=1. \label{eq:EbalJ}
\end{align}
The energy spent to travel from $d_i$ to $o_j$ is $E_{ij}^k$. Moreover, $\bar{E}_\mathrm{con}$ is the consumption per unit distance of the vehicles, $d^\mathrm{fp}_{ij}$ is the distance of the fastest path from $d_i$ to $o_j$, and $\Delta d_{ijc}^\mathrm{go2S}$ is the additional distance driven to pass by station~$c$.
Then, $e^k_{j}$ is the energy of vehicle $k$ at $d_j$.
Finally, we set initial and final battery state of charge for every vehicle and we bound its state of energy to be within its capacity $E_\mathrm{b}^\mathrm{max}$ as
\begin{equation}
	0 \leq e^k_{j} \leq E_ \mathrm{b}^\mathrm{max} \; \;\forall j \in \cI^+,\;\forall k\in\cK,
\end{equation}
\begin{equation}\label{eq:boundTime}
	e^k_{0}= e^k_{I+1} = E_\mathrm{b}^0\;\;\forall k\in\cK.
\end{equation}
Note that $E_\mathrm{b}^\mathrm{max}$ represents the usable battery, which we assume to be lower than the nominal battery capacity so that the operation of the fleet is not influenced by battery degradation. We highlight that the assumption is in order to decrease battery degradation and enforce a safe lower bound on the battery state of charge. 

\subsection{Objective}
In this paper, the multi-objective function which takes into account not only revenues generated by serving travel requests and operational costs (revenues) generated by performing (dis)charging activities, but also operational costs due to battery degradation.
Formulated as a cost-minimization function, the objective is then
\begin{equation}\label{eq:obj}
	J=  J_\mathrm{trav} + J_\mathrm{elec} +  J_\mathrm{batt}.
\end{equation}
The first term represents the revenue generated by serving travel requests, the second represents the cost and revenue caused by charging and discharging, and the last term is the cost due to battery degradation. In particular the first two terms can be expressed in linear form by
\begin{equation}\label{eq:obj}
	J_\mathrm{trav} =  - \sum_{i \in \cI} b_\mathrm{r}^i \cdot p_i,
\end{equation}
\begin{equation}\label{eq:obj2}
	J_\mathrm{elec} =  \sum_{i,j\in \cI} p_{ij}^\mathrm{el} \cdot \sum_{k \in \cK} \sum_{c \in \cC} C^k_{ijc},
\end{equation}
where $p_{ij}^\mathrm{el}$ is the average price of electricity during transition $ij$, the binary variable $b_\mathrm{r}^i$ indicates whether request $i$ is served or rejected, and $p_i$ is the revenue generated by accepting such request.
In the next section we will detail $J_\mathrm{batt}$.
\subsection{Battery Degradation}
During the lifetime of a vehicle, its battery deteriorates due to irreversible electro-chemical reactions, known as battery aging. The predominant part of this phenomenon is called "cyclic ageing", and occurs during the charging and discharging of the battery. The normalized battery capacity degradation, similar to~\cite{RathMedinaEtAl2023}, is defined as
\begin{equation}\label{eq:degr}
	\Delta \tilde{E_\mathrm{b}} = \frac{\kappa}{\sqrt{V_\mathrm{ch}}} \cdot \sqrt{Q},
\end{equation}
\begin{equation}
	\kappa = b_1 \cdot (\phi_z - b_2)^2 + b_3 \cdot \Delta z
	+ b_4 \cdot C_\mathrm{rate}^\mathrm{ch} + b_5 \cdot C_\mathrm{rate}^\mathrm{dch} + b_6,
\end{equation}
where $V_\mathrm{ch}$ is the charging voltage, $Q$ is the energy flowing through the battery, the average SoC is $\phi_z$, while $\Delta z$ is the Depth of Discharge (DoD), $C_\mathrm{rate}^\mathrm{ch}$ and $C_\mathrm{rate}^\mathrm{dch}$ are the \emph{C-rates} during a full charge and discharge cycles, respectively. 
The values of the battery $b_1$ to $b_6$ are battery-specific, depend on the type of battery, and can be found in Table~\ref{tablevalues}. Following the same reasoning of~\cite{RathMedinaEtAl2023}, the initial and final battery state of charge are fixed, so we assume $\phi_z,\Delta z$ are not strongly influenced by the operation of the fleet. Assuming $C_\mathrm{rate}^\mathrm{ch}= C_\mathrm{rate}^\mathrm{dch} = C_\mathrm{rate}$ to be constant, $\kappa$ does not depend on the optimization variables of the problem. However,~\eqref{eq:degr} is non-linear, meaning that the operational cost due to battery degradation would not be constant throughout the lifetime of each vehicle. Nevertheless, from an operational point of view, the important key factor to consider is the overall cost throughout the whole battery lifetime, i.e., the average cost per day. Thus, we approximate~\eqref{eq:degr} with a linear function so that the average cost per day over the whole lifetime is captured.

\begin{figure}[t]
	\centering
	\includegraphics[width=\columnwidth]{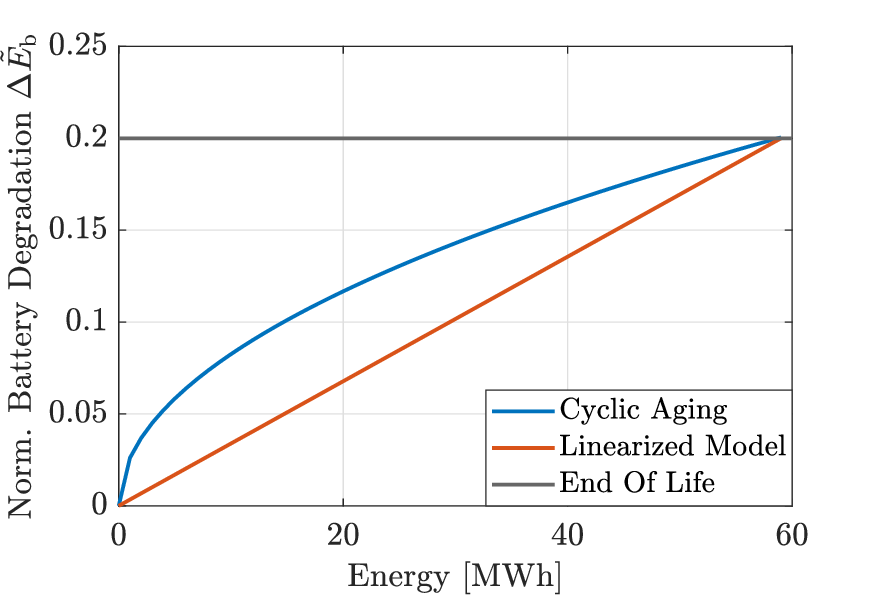}
	\caption{Normalized battery degradation model and linearized version as a function of the overall energy flown through a $\unit[40]{kWh}$ battery.} 
	\label{fig:deg}
\end{figure}

We frame the approximated cost function for the whole fleet as
\begin{equation}\label{eq:jbattlin}
	{J}_\mathrm{batt} =  \frac{p_\mathrm{batt}}{Q^\mathrm{eol}} \cdot \sum_{i,j,c,k}\abs{C^k_{ijc}}
\end{equation}  
\begin{equation}\label{eq:jbattlin}
Q^\mathrm{eol} = \left(
\frac{\Delta E^\mathrm{eol} }{\kappa}\right)^2 \cdot V_\mathrm{ch},
\end{equation}
where $p_\mathrm{batt}$ is the cost of the battery, including the end of life (EoL) value, and $Q^{\mathrm{eol}}$ is the overall (dis)charging that a battery can withstand before EoL, which is obtained by imposing $ \Delta \tilde{E_\mathrm{b}} = \Delta E_\mathrm{b}^\mathrm{eol}$, the normalized battery capacity degradation before EoL for automotive applications, in~\eqref{eq:degr}, as shown in Fig.~\ref{fig:deg}. 

\subsection{Problem Definition}
We define the minimum (negative) profit problem for an electric fleet of vehicles for MaaS as follows:

\begin{prob}[MaaS Problem]\label{prob:main}
Given a set of transportation requests $\cI$, the operations maximizing the total profit of the electric fleet result from
	\begin{equation*}
		\begin{aligned}
			\min\; &J\\
			\mathrm{s.t. }\; &\eqref{eq:boundX}-\eqref{eq:obj2},\eqref{eq:jbattlin}.
		\end{aligned}
	\end{equation*}
\end{prob}
Problem~\ref{prob:main} is a multi-objective mixed integer linear program that can be solved with global optimality guarantees, or at least within a guaranteed optimality gap above such optimum.

\subsection{Discussion}
A few comments are in order. First,  we assume the fleet size is small enough to not influence electricity price and road congestion. In other words, the routing of the fleet does not affect neither travel time nor electricity price.
Second, we assume both electricity price and travel requests to be known in advance during the entire day. In fact, Problem~\ref{prob:main}  is not suited for an online implementation, but it allows to compute a cost-wise lower bound of an online implementation. Last, it is possible to obtain an online implementation of this model by leveraging a receding horizon approach, i.e., model predictive control, where it is possible to account for new appearing travel requests.

%% file: Sections/CaseStudy.tex
\section{Results}\label{sec:Res}
\begin{figure}[t]
	\centering
	\includegraphics[width=0.95\columnwidth]{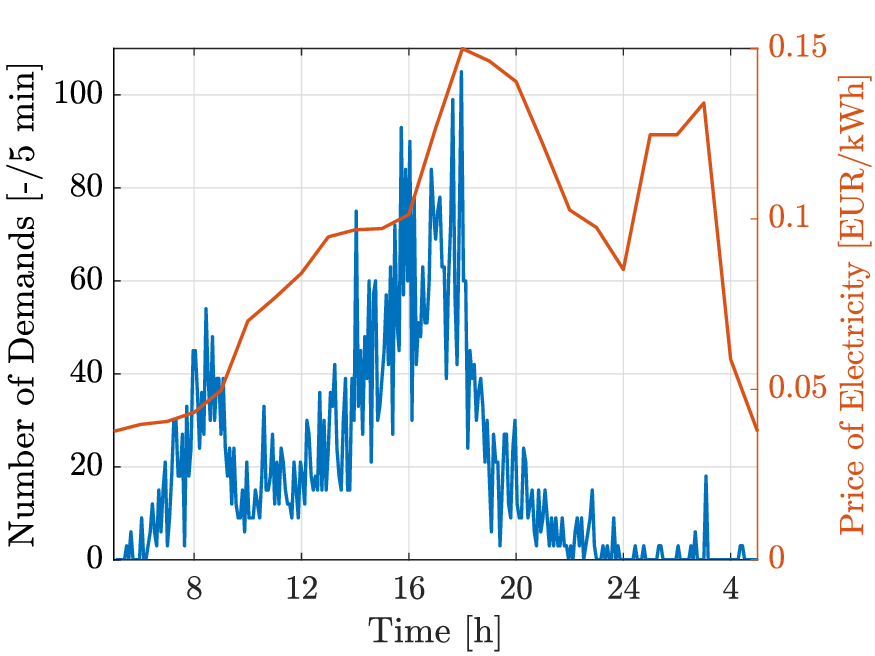}
	\caption{The figure shows the number of demands during a day, and the recorded energy price in the Netherlands during November 1st, 2023 (courtesy of nordpoolgroup.com).} 
	\label{fig:DemPrice}
\end{figure}

In this section, we show a MaaS system in terms of operational strategies and costs in the cases of no battery degradation and no V2G activities. Then, we show the operation of the fleet in terms of overall exchanged energy w.r.t.\ the end of life battery price $p_\mathrm{batt}$.
We present a case study conducted in the city of Eindhoven, The Netherlands, using data generated by Albatross, A Learning BAsed TRansportation Oriented Simulation System, \cite{ArentzeTimmermans2004,RasouliKimEtAl2018}. Albatross is a state-of-the-art activity-based model from the computational process model family, employing a series of Chi-square Automatic Interaction Detection (CHAID) decision trees at its core, to simulate activity-travel schedules. Albatross is household-based meaning that up to two household heads are included in the household. Initially, a synthetic population is simulated for Eindhoven using the iterative proportional fitting method. Subsequently, the synthetic population is used as input for the activity-travel model. Albatross simulates the activity-travel schedules for this population for a complete day. The model simulates leave times, travel times, start times of the activity, end times of the activity, origin, destination (at Postal Code level 4, i.e., the first 4 digits of the Dutch postal code), activity type and mode choice. Fig.~\ref{fig:DemPrice} shows the number of travel requests and the price of electricity during the day, for the Dutch city of Eindhoven: Our analysis focuses on approximately 2000  randomly selected travel requests, served by a fleet of 70 Nissan Leaf 2022, comparing the performance of the MaaS system for a varying price of battery $p_\mathrm{batt}$. In particular we show the results in two extreme cases where battery degradation is neglected and where V2G is not allowed. 

\begin{table}[t]
	\caption{Parameter Values}
	\label{tablevalues}
	\begin{center}
		\begin{tabular}{lcl}
			\toprule
			Parameter & Value &  Unit \\
			\midrule
	$p_\mathrm{batt}$ & $4000$ & EUR\\
	$E_\mathrm{batt}$ & $40$ & kW\\               				
	$P_\mathrm{ch}$ & $22$ & kW\\
	$\bar{E}_\mathrm{con}$ & $0.15$ & kWh/km\\
	$\Delta E_\mathrm{b}^\mathrm{eol}$ & $0.2$ & - \\ 
	$b_1$ & $-2.87 \cdot 10^{-4}$ & -\\
	$b_2$ & $3.352\cdot 10^{-2}$ & -\\
	$b_3$ & $3.8\cdot 10^{-3}$ & -\\
	$b_4$ & $3.578\cdot 10^{-5}$ & -\\
	$b_5$ & $2.274\cdot 10^{-4}$ & -\\
	$b_6$ & $1.02\cdot 10^{-2}$ & -\\
			\bottomrule
		\end{tabular}
	\end{center}
\end{table} 

\begin{figure}[t!]	
	\centering
	\includegraphics[trim={0 0 0 0},clip,width=0.99\linewidth]{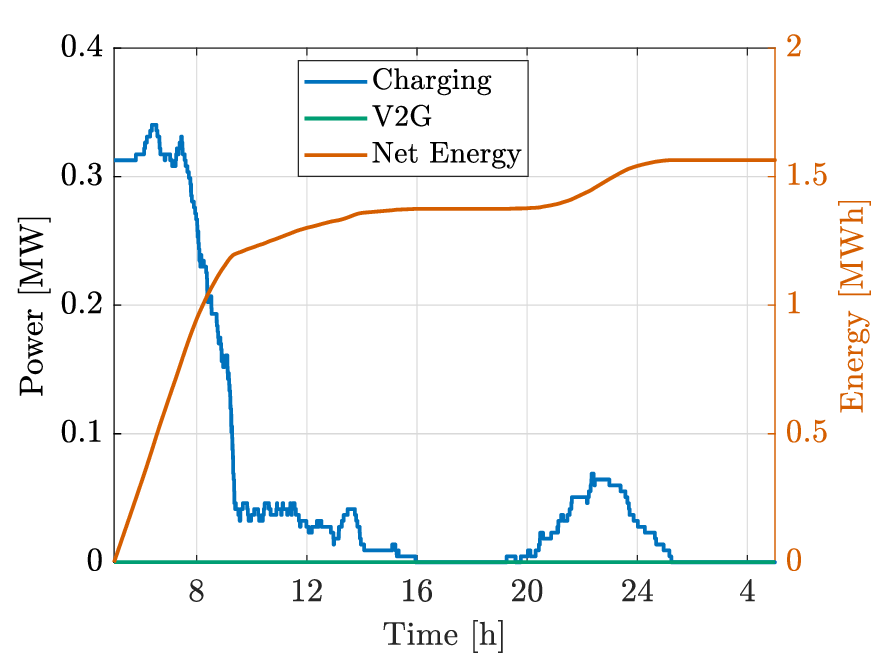}
	\includegraphics[trim={0 0 0 0},clip,width=0.99\linewidth]{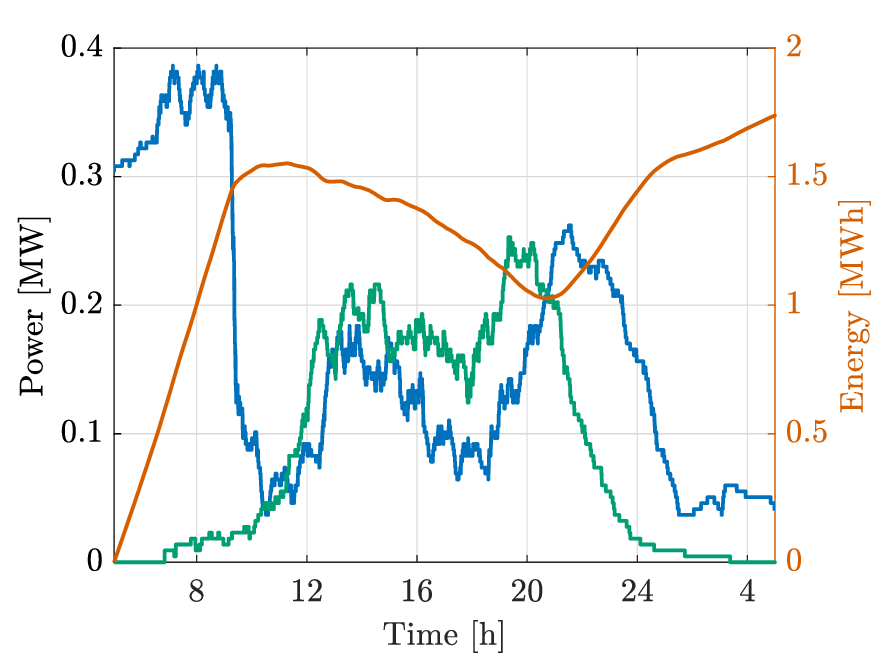}
	\caption{Charging, discharging activities and net energy withdrawn from the grid during the day. In the top figure the fleet is not allowed to perform V2G, while in the bottom figure the fleet can perform V2G. Fleet composed of $70$ Nissan Leaf. Power of the chargers of $\unit[22]{kW}$.}
	\label{fig:V2G}
\end{figure}

\begin{table}[t]
	\caption{Daily Operational Costs and Revenues to operate a fleet of 70 Nissan Leaf.}
	\label{table}
	\begin{center}
		\begin{tabular}{lclc}
			\toprule
			Metrics & Unaware V2G &  No V2G & Unit \\
			\midrule
			Travel Requests Revenue & $21900$ & $21900$ & EUR/day\\
			Charged Energy & $6050$ & $1780$ & kWh\\
			Charging Cost & $610$ & $165$ & EUR/day\\
			Discharged Energy & $4190$ & $0$ & kWh\\
			Discharging Revenue & $750$ & $0$ & EUR/day \\
			Battery Degradation Cost &$690$ & $120$& EUR/day\\
			Profit &$21350$ & $21615$ & EUR/day\\	
			Avg. Battery Lifetime & $405$ & $2360$ & day\\ 
			\bottomrule
		\end{tabular}
	\end{center}
\end{table} 

\begin{table}[t]
	\caption{Daily Operational Costs and Revenues  per vehicle to operate a fleet of Nissan Leaf with aware and unaware battery degradation operations. Battery price \mbox{$p_\mathrm{b}=\unit[4000]{EUR}$}, with battery capacity of $\unit[40]{kWh}$.}
	\label{table2}
	\begin{center}
		\begin{tabular}{lclc}
			\toprule
			Metrics & Unaware V2G &  Aware V2G & Unit \\
			\midrule
			Travel Requests Revenue & $354$ & $354$ & [EUR/day]\\
			Charged Energy & $84$ & $47$ & [kWh]\\
			Charging Cost & $8.5$ & $3.5$ & [EUR/day]\\
			Discharged Energy & $57$ & $12$ & [kWh]\\
			Discharging Revenue & $10.3$ & $3.3$ & [EUR/day] \\
			Battery Degradation Cost &$6.5$ & $4$& [EUR/day]\\
			Profit &$346.2$ & $349.8$ & [EUR/day]\\	
			Battery Lifetime & $418$ & $1000$ & [day]\\ 
			\bottomrule
		\end{tabular}
	\end{center}
\end{table} 
\subsection{Case Study of Eindhoven, The Netherlands}
\begin{figure}[!t]
	\centering
	\includegraphics[width=\columnwidth]{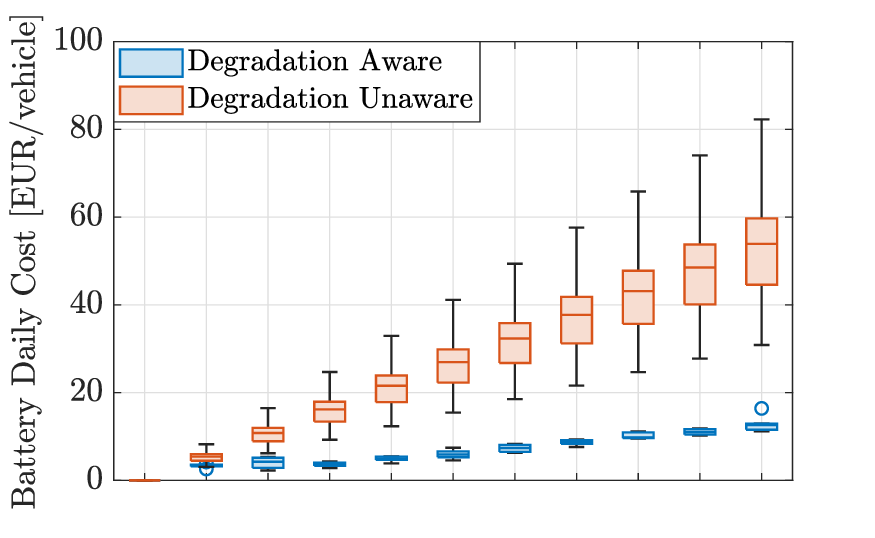}
	\includegraphics[width=\columnwidth]{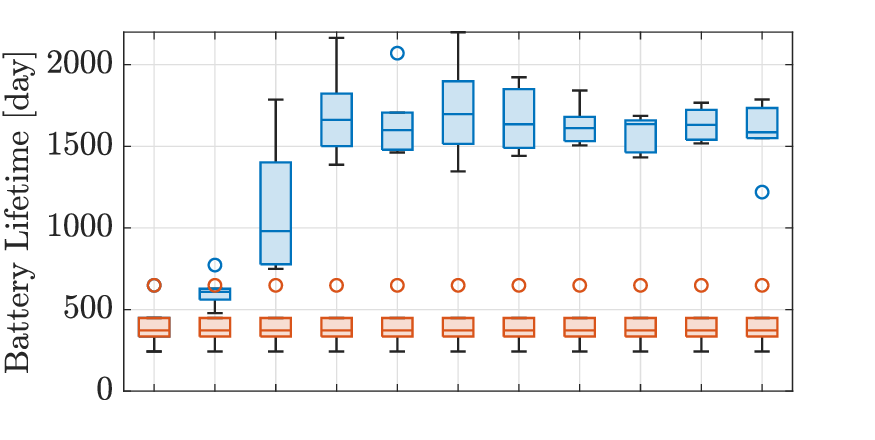}
	\includegraphics[width=\columnwidth]{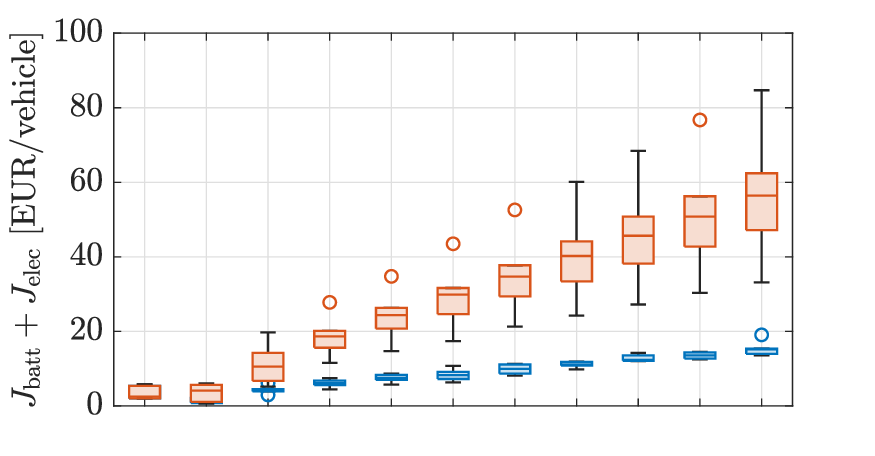}
	\includegraphics[width=\columnwidth]{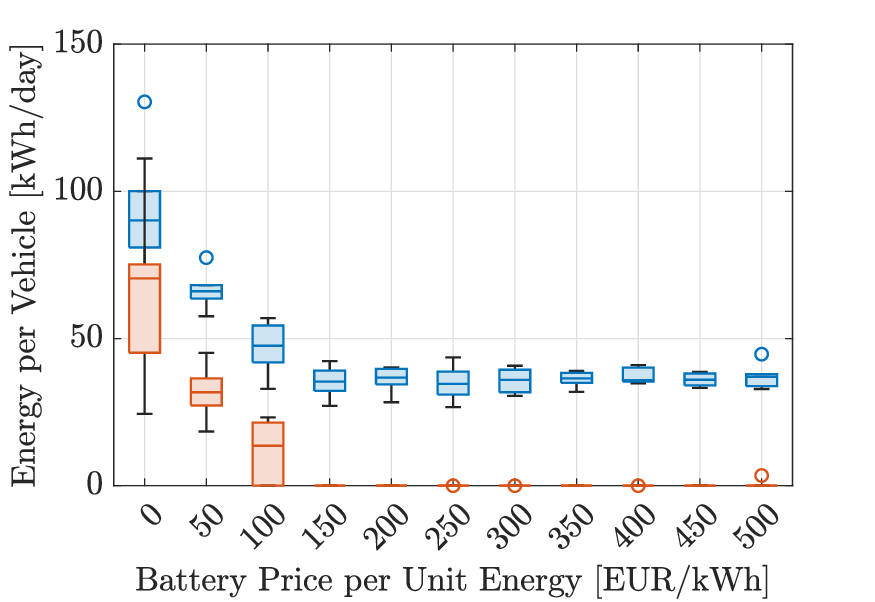}
	\caption{Costs due to degradation per day per vehicle, battery lifetime, objective excluding serving travel requests, and optimal average charged  and discharged energy per vehicle as a function of the battery end of life price per unit energy. Battery capacity of Nissan Leaf of $\unit[40]{kWh}$.} 
	\label{fig:pareto}
\end{figure}

In this section, we analyze the results obtained by solving Problem~\ref{prob:main} for the case study of Eindhoven. Fig~\ref{fig:V2G} shows the power withdrawn from and injected to the grid, as well as the overall energy consumed by the electric fleet during one day of operation. The figure on the top shows the first case where battery degradation is very high, i.e., V2G activities are not allowed. The bottom figure, on the contrary, shows the case were battery degradation is neglected. Table~\ref{table} shows the overall profit generated in the two cases, with battery degradation estimated with~\eqref{eq:degr}, after solving Problem~\ref{prob:main}.
Interestingly, we see that the revenues generated by serving the travel requests are not influenced by the V2G activities.  The results show that considering battery degradation is crucial, and that not considering it may lead to a counter-productive result. 
Given that the revenues generated by travel requests are not influenced by V2G activities, i.e., the vehicles have extra time to perform (dis)charging activities, in what follows we repeat similar simulations, but with a slightly smaller fleet, 60 Nissan Leaf. {Fig.~\ref{fig:pareto}} shows several metrics as a function of the battery price $p_\mathrm{b}$ normalized by the capacity. The result confirms the previous findings: it is crucial to include battery degradation in the objective when dealing with V2G activities. In fact, the result with $p_\mathrm{b}= 0 $, which is the equivalent of the V2G case in Fig.~\ref{fig:V2G}, shows that the charging and discharging activities are more intense compared to the other cases with $p_\mathrm{b} > 0 $, and that they become unprofitable for normalized battery price above $\unit[100]{EUR/kWh}$, equivalent to $p_\mathrm{b}=\unit[4000]{EUR}$ for a Nissan Leaf with battery capacity of $\unit[40]{kWh}$.
Table~\ref{table2} shows the results for the unaware V2G case against aware V2G with normalized battery price of $\unit[100]{EUR/kWH}$. We highlight that, compared to the previous simulation with more vehicles, the lower number of vehicles leads to a higher revenue per vehicle, a lower V2G activity, a lower degradation and a higher rejection rate of the travel requests, but shows that the profitability results are consistent over different scenarios and over different ratios between travel requests and vehicles. This confirms that V2G activities are done when there are fewer travel requests to be served. Then, the battery lifetime in the last row was estimated leveraging~\eqref{eq:degr}, confirming that the approximation used to maintain the linear structure of Problem~\ref{prob:main} yields to consistent results. 
Last, we emphasize that the threshold up to which V2G activities are unprofitable are strongly influenced by the average energy price, which in our case, is lower compared to the average consumer price. 

%% file: Sections/Conclusions.tex
\section{Conclusions}\label{Sec:Concl}
In this paper, we presented a modeling and optimization framework that allows to analyze the optimal operation of a fleet of electric vehicles that provides Mobility-as-a-service, whilst performing V2G activities.
In particular, we captured the trade-off between intensity of charging and discharging activities of the fleet, from which it is possible to generate revenues by trading energy, and the operational cost due to battery degradation, which is mainly caused by charging and discharging activities. We formulated the problem as a multi-objective mixed integer linear program. The results of the simulations performed in Eindhoven, The Netherlands, showed that when performing V2G, it is crucial to take battery degradation into account  due to the high operational cost, which can overweigh the additional revenue generated by trading energy, leading to an overall lower profit. In the future, we would like to have an online implementation of the problem leveraging stochastic model predictive control to take into account uncertainty of travel requests and energy prices, and measure the optimality gap w.r.t.\ the offline version.